\documentstyle[11pt]{article}

\newcommand{\half}{{\scriptstyle{\frac{1}{2}}}}
\begin{document}
\begin{titlepage}
\begin{flushright}
{\small DOE/ER/40717-28 \\
May 1996}
\end{flushright}

\vspace*{14mm}
\begin{center}
               {\Large\bf Response to Tarrach's ``Mode Dependent \\
\vspace*{4mm}
                  Field Renormalization and Triviality''}

\vspace*{18mm}
{\Large  M. Consoli}
\vspace*{5mm}\\
{\large
Istituto Nazionale di Fisica Nucleare, Sezione di Catania \\
Corso Italia 57, 95129 Catania, Italy}
\vspace*{5mm}\\
and 
\vspace*{5mm}\\
{\Large P. M. Stevenson}
\vspace*{5mm}\\
{\large T.~W.~Bonner Laboratory, Physics Department \\
Rice University, Houston, TX 77251, USA}
\vspace{16mm}\\
{\bf Abstract:}
\end{center}
 
\par 
We respond to Tarrach's criticisms of our work on $\lambda \Phi^4$ 
theory.  Tarrach does not discuss the same renormalization procedure 
that we do.  He also relies on results from perturbation theory that 
are not valid.  There is no ``infrared divergence'' or unphysical 
behaviour associated with the zero-momentum limit of our effective action.
  
\end{titlepage}
 
\setcounter{page}{1}

     In a recent paper Tarrach \cite{tarrach} has criticized our work 
on $(\lambda \Phi^4)_4$ theory in which we obtain a ``trivial'' but not 
entirely trivial continuum limit \cite{cs}.  However, ({\bf 1}) Tarrach 
does not consider the same renormalization procedure that we do, and 
thus his ``main result'' (Eq. (23)) has no relevance to our proposal; 
({\bf 2}) his discussion assumes results from perturbation theory that 
are not valid; and ({\bf 3}) his implication that there is something 
physically pathological about the zero-momentum limit of our effective 
action is not true.

{\bf 1.}
     Although comparison is somewhat obscured by Tarrach's very different 
terminology and notation, there is an easy way to see that he is discussing 
a quite different renormalization procedure from ours.  We both consider 
a re-scaling of the zero-momentum mode of the field, and hence of its 
vacuum value $v$, but Tarrach's is different from ours.  In our work the 
key requirement is that the combination $\lambda_B v_B^2$, governing the 
physical mass, should be finite.  In our notation $\lambda_B$ is the 
bare coupling constant, which tends to zero like $1/\ln ({\rm cutoff})$, 
and the finite, physical $v$ is related to the bare field by
\begin{equation}
          v_B = Z_{\phi}^{1/2} v
\end{equation}
with $Z_{\phi} \sim \ln ({\rm cutoff})$, so that $1/Z_{\phi}$ {\it scales 
like} $\lambda$.  In Tarrach's paper the corresponding equation is in the 
last line of Eq. (20):
\begin{equation}
{\mbox{{\rm ``}}} v_R = Z_A^{-1/2} A {\mbox{{\rm ''}}},
\end{equation}
where ``$v_R$'' is essentially our $v_B$ (it is ``$Z_R^{-1/2} v_B$'' with 
``$Z_R$'' $\sim 1$) and ``$A$'' is the finite quantity (our $v$).  Thus, 
Tarrach's ``$Z_A$'' is $1/Z_{\phi}$.  However, {\it it does not scale like} 
$\lambda$:  in his continuum limit (``$\tau \to 0$''), it scales as 
$\mid \! \ln \tau \! \mid^{-1/2}$ (his Eq. (21)) while $\lambda$ scales 
as $\mid \! \ln \tau \! \mid^{-1}$ (his Eq. (19)).  Thus, Tarrach's 
renormalization is not ours.  The fact that he finds no surviving mass 
term in his renormalized effective action (Eq. (23)) is unsurprising, and 
has no bearing on our work.  

      Though, for reasons to be explained below, we do not accept Tarrach's 
initial premise, Eq. (17), it might be instructive to point out that he 
could have produced a more accurate caricature of our picture by replacing 
his postulated Eq. (18) with 
\begin{equation}
          a \sim \tau^{1/2} \mid \! \ln \tau \! \mid^{-1/6} L.
\end{equation}
This would yield our re-scaling for $v$ and also an ``$m_R$'' that is 
finite in physical units.  Superficially, it leads to an effective potential 
that is of order $\ln ({\rm cutoff})$, but in our picture, as originally 
in Ref. \cite{st}, this is remedied by a cancellation.  This cancellation 
is simply the fact that a function made up of a log-divergent $\phi^4$ term 
and a finite $\phi^4 \ln \phi^2$ term can always be re-written as 
$\phi^4 ( \ln \phi^2/v^2 - \half )$, with the divergence absorbed into 
the vacuum value $v$.

{\bf 2.} 
     Tarrach's starting point, his Eq. (17), relies on results from 
renormalization-group-improved perturbation theory (RGIPT).  He claims 
that these results are ``very solidly founded, because RGIPT is, at low 
energies, and {\it because of triviality} [our italics], very reliable.''   
This is a common misconception: It falsely assumes that a small (or 
vanishingly small) renormalized coupling is a sufficient condition for 
RGIPT to work.  In fact, the traditional approach and ``triviality'' are 
inherently contradictory about the continuum limit; the former begins by 
postulating a finite, {\it non-zero} renormalized coupling constant, and 
``triviality'' says that there can be no such thing.  

     In \cite{trivpert} we discuss exactly what goes wrong with RGIPT: Its 
re-summation of leading logs tries to re-sum a geometric series that is 
inevitably {\it divergent} when one tries to take the continuum limit.  
Our not-entirely-trivial continuum limit arises precisely where the 
leading-log series becomes $1-1+1-\ldots$, which RGIPT assumes will re-sum 
to $1/(1+1)=1/2$.  There are instances in physics where such an illegal 
re-summation happens to give the right answer --- but this is not one of them.

     Tarrach's Eq. (17) assumes, based on perturbation theory, that 
spontaneous symmetry breaking (SSB) in lattice $(\lambda\Phi^4)_4$ theory 
corresponds to a {\it second-order} phase transition.  This is not true 
in our picture, and recent lattice data \cite{agodi} strongly supports 
our claim.  {\it A priori}, for a given value of the bare coupling constant, 
$\lambda_B$, one can define two distinct critical values of the 
bare-mass-squared parameter $r \equiv m_B^2$; one, $r_{\rm PhT}$, is where 
the phase transition actually occurs; the other, $r_{\rm CSI}$, is where the 
mass gap of the symmetric phase becomes exactly zero (the ``classically 
scale-invariant'' (CSI) case).  If these two values exactly coincide then 
the transition is second order.  If that were so, then a continuum limit 
could be obtained for {\it any} $\lambda_B$ by taking the limit $\tau \to 0$, 
where $\tau= |1- {{r}\over{r_{\rm CSI} } }|$), since the physical correlation 
length would then diverge in units of the lattice spacing.

     However, to find out whether $r_{\rm CSI}$ and $r_{\rm PhT}$ coincide, 
one must explore the effective potential of the theory.  As discussed in 
our papers \cite{cs}, in any approximation consistent with ``triviality'' 
--- i.e., one in which the shifted field $h(x)=\Phi(x)-\langle\Phi\rangle$ 
is effectively governed by a quadratic Hamiltonian, with its propagator 
determined by solving exactly a non-perturbative gap equation --- the 
massless theory at $r=r_{\rm CSI}$ lies within the broken phase; i.e., 
$r_{\rm CSI}$ is more negative than $r_{\rm PhT}$.  Our approach predicts 
that the exact form of the effective potential in the CSI case is 
$\phi^4(\ln \phi^2/v^2 - \half)$, and this has been confirmed to great 
accuracy by lattice simulations \cite{agodi}.  

      Since $r_{\rm CSI}$ and $r_{\rm PhT}$ differ, the phase transition is 
{\rm first-order}.  In order to obtain a continuum limit, one needs the 
physical correlation length $\xi_h$ of the {\it broken} phase to be infinite 
in units of the lattice spacing.  In other words, the mass $m_h \sim 1/\xi_h$ 
of the fluctuations about the SSB vacuum must be much, much less than the 
cutoff.  As discussed in our papers \cite{cs,agodi}, this requires 
$\lambda_B$ to tend to zero like $1/\ln({\mbox{{\rm cutoff}}})$ \cite{cs}.  

      With such a $\lambda_B$, although $r_{\rm CSI}$ and $r_{\rm PhT}$ differ, 
they differ -- even in physical units -- only by an infinitesimal amount: 
each is negative and huge, of order (cutoff)$^2$, while their difference is 
infinitesimal, of order $1/\ln(\mbox{{\rm cutoff}})$.  However, all the 
interesting physics occurs over such an infinitesimal range of $r$ around 
$r_{\rm PhT}$.  This is because 
such tiny variations in $r$ cause {\it finite} changes (i) in the particle 
mass of the broken vacuum, (ii) in the energy-density difference between the 
two phases, and (iii) in the barrier between them.  The problem with the 
conventional approach is that it looks at the phase transition on too coarse 
a scale --- making finite variations in $r$.  Viewed on that scale the 
transition appears indistinguishable from a second-order transition and the 
not-entirely-trivial physics is not seen.  

{\bf 3.}
     Tarrach also alleges that our effective action is ``infrared 
divergent.''  It is not clear what he means by this.  There is, of course, 
the usual infinite-volume factor in the relation between the effective 
action and the effective potential.  In a derivative expansion of the 
effective action the term with no derivatives is 
$- \int \! d^4 x \, V_{{\rm eff}}(\Phi(x))$, so that if $\Phi(x) = \phi =$ 
constant one gets $ - (\int \! d^4 x) \, V_{{\rm eff}}(\phi)$ (see, eg. 
\cite{cw}).  
Physically, this is natural --- the energy diverges with the volume if the 
energy density is finite --- but it is rather improper mathematics.  
Tarrach objects to having a {\it constant} source, and hence a constant 
$\phi$, $\neq v$, insisting that all sources should fall off to zero at 
infinity.  However, that is only one way of regularizing.  More conveniently 
the theory can be formulated in finite volume with periodic boundary 
conditions; there is then no problem with considering a source that is 
constant over this volume.  An excellent treatment of our picture in a 
finite-volume formalism has been given by Ritschel \cite{ritschel}.  
This issue has nothing to do with our non-traditional ultraviolet 
renormalization.  

     It is true that our renormalized effective action is discontinuous 
at zero momentum, in that the renormalized proper $n$-point functions 
($n \ge 3$) are zero at finite momentum, but are non-zero at zero momentum.  
[Our renormalized 2-point function, however, has {\it no} discontinuity at 
zero momentum; our field renormalization is precisely what is needed to 
ensure this.]  However, this discontinuity could never be directly revealed 
experimentally, because scattering experiments with exactly zero-momentum 
particles are inherently impossible.  Moreover, $S$-matrix elements are 
more directly related, not to the proper Green's functions generated by the 
effective action, but to the {\it full} Green's functions.  The latter are 
inherently singular at $p=0$, whenever there is SSB, because they contain 
disconnected pieces proportional to $\delta^{(4)}(p)$.  Smoothness at 
$p \to 0$ is not to be expected since the underlying phenomenon is Bose 
condensation.  Macroscopic occupation of the $p=0$ mode gives it a unique 
status, making it entirely natural that it requires its own special 
re-scaling.  

\vspace*{8mm}
\begin{center}
{\bf Acknowledgements}
\end{center}

We thank Rolf Tarrach for correspondence. 

This work was supported in part by the U.S. 
Department of Energy under Grant No. DE-FG05-92ER40717.


\end{document}